\PassOptionsToPackage{unicode}{hyperref}
\PassOptionsToPackage{hyphens}{url}
\PassOptionsToPackage{dvipsnames,svgnames,x11names}{xcolor}
\documentclass[
  11pt,
  a4paper,
]{article}
\usepackage{amsmath,amssymb}
\usepackage{setspace}
\usepackage{iftex}
\ifPDFTeX
  \usepackage[T1]{fontenc}
  \usepackage[utf8]{inputenc}
  \usepackage{textcomp} 
\else 
  \usepackage{unicode-math} 
  \defaultfontfeatures{Scale=MatchLowercase}
  \defaultfontfeatures[\rmfamily]{Ligatures=TeX,Scale=1}
\fi
\usepackage[]{mathpazo}
\ifPDFTeX\else
\fi
\IfFileExists{upquote.sty}{\usepackage{upquote}}{}
\IfFileExists{microtype.sty}{
  \usepackage[]{microtype}
  \UseMicrotypeSet[protrusion]{basicmath} 
}{}
\makeatletter
\@ifundefined{KOMAClassName}{
  \IfFileExists{parskip.sty}{%
    \usepackage{parskip}
  }{
    \setlength{\parindent}{0pt}
    \setlength{\parskip}{6pt plus 2pt minus 1pt}}
}{
  \KOMAoptions{parskip=half}}
\makeatother
\usepackage{xcolor}
\usepackage[margin=1in]{geometry}
\usepackage{longtable,booktabs,array}
\usepackage{calc} 
\usepackage{etoolbox}
\makeatletter
\patchcmd\longtable{\par}{\if@noskipsec\mbox{}\fi\par}{}{}
\makeatother
\IfFileExists{footnotehyper.sty}{\usepackage{footnotehyper}}{\usepackage{footnote}}
\makesavenoteenv{longtable}
\usepackage{graphicx}
\makeatletter
\def\maxwidth{\ifdim\Gin@nat@width>\linewidth\linewidth\else\Gin@nat@width\fi}
\def\maxheight{\ifdim\Gin@nat@height>\textheight\textheight\else\Gin@nat@height\fi}
\makeatother
\setkeys{Gin}{width=\maxwidth,height=\maxheight,keepaspectratio}
\makeatletter
\def\fps@figure{htbp}
\makeatother
\NewDocumentCommand\citeproctext{}{}
\NewDocumentCommand\citeproc{mm}{%
  \begingroup\def\citeproctext{#2}\cite{#1}\endgroup}
\makeatletter
 \let\@cite@ofmt\@firstofone
 \def\@biblabel#1{}
 \def\@cite#1#2{{#1\if@tempswa , #2\fi}}
\makeatother
\newlength{\cslhangindent}
\newlength{\csllabelwidth}
\newenvironment{CSLReferences}[2] 
 {\begin{list}{}{%
  \setlength{\itemindent}{0pt}
  \setlength{\leftmargin}{0pt}
  \setlength{\parsep}{0pt}
  \ifodd #1
   \setlength{\leftmargin}{\cslhangindent}
   \setlength{\itemindent}{-1\cslhangindent}
  \fi
  \setlength{\itemsep}{#2\baselineskip}}}
 {\end{list}}
\usepackage{calc}

\usepackage{amsmath} 

\usepackage{lscape}
\newcommand{\blandscape}{\begin{landscape}}
\newcommand{\elandscape}{\end{landscape}}

\newcommand{\bleft}{\begin{flushleft}}
\newcommand{\eleft}{\end{flushleft}}

\ifLuaTeX
  \usepackage{selnolig}  
\fi
\usepackage{bookmark}
\IfFileExists{xurl.sty}{\usepackage{xurl}}{} 
\hypersetup{
  pdftitle={Spatial occupancy models for data collected on stream networks},
  pdfauthor={Olivier Gimenez1*},
  colorlinks=true,
  linkcolor={RoyalBlue},
  filecolor={Maroon},
  citecolor={Blue},
  urlcolor={RoyalBlue},
  pdfcreator={LaTeX via pandoc}}

\title{Spatial occupancy models for data collected on stream networks}
\author{Olivier Gimenez\textsuperscript{1}*}
\date{2024-09-10}

\begin{document}
\maketitle

\setstretch{2}
\small

\textsuperscript{1} CEFE, Univ Montpellier, CNRS, EPHE, IRD, Montpellier, France

\texttt{*} Corresponding author: \href{mailto:olivier.gimenez@cefe.cnrs.fr}{\nolinkurl{olivier.gimenez@cefe.cnrs.fr}}

\normalsize

\vspace{1cm}
\hrule

To effectively monitor biodiversity in streams and rivers, we need to quantify species distribution accurately. Occupancy models are useful for distinguishing between the non-detection of a species and its actual absence. While these models can account for spatial autocorrelation, they are not suited for streams and rivers due to their unique network spatial structure. Here, I propose spatial occupancy models specifically designed for data collected on stream and river networks. I present the statistical developments and illustrate their application using data on a semi-aquatic mammal. Overall, spatial stream network occupancy models offer a robust method for assessing biodiversity in freshwater ecosystems.

\vspace{3mm}
\hrule
\vspace{5mm}


\bleft
\newpage

\section{Introduction}\label{introduction}

Streams and rivers provide essential habitats for numerous species of animals and plants (\citeproc{ref-Reid2019}{Reid et al. 2019}). The ecological health of these freshwater ecosystems is paramount not only for the biodiversity they harbor but also for the ecosystem services they provide, which are indispensable to both wildlife and human populations (\citeproc{ref-V2022}{Vári et al. 2021}). However, human activities are altering the natural conditions of streams, rivers, and their associated riparian habitats, jeopardizing the persistence of these ecosystems (\citeproc{ref-A2020}{Albert et al. 2020}).

In this context, species distribution models (SDMs) are essential tools in understanding and preserving biodiversity (\citeproc{ref-Elith2009}{Elith and Leathwick 2009}). SDMs predict the distribution of species, helping scientists and conservationists in identifying critical habitats and biodiversity hotspots. Additionally, SDMs inform strategies aimed at mitigating the impacts of climate and land-use changes, managing invasive species, and enhancing habitat connectivity in freshwater ecosystems (\citeproc{ref-D2015}{Domisch et al. 2015}).

SDMs are influenced by two main issues: imperfect detection and spatial autocorrelation (\citeproc{ref-GK2018}{Guélat and Kéry 2018}). First, imperfect detection occurs when a species present in a given area is not detected during surveys due to factors such as observer experience, species behavior, and environmental conditions. Ignoring imperfect detection can lead to biased estimates of species distribution and flawed inferences about the relationship between species presence and environmental factors (e.g., \citeproc{ref-Lahoz2014}{Lahoz-Monfort et al. 2014}). This can misinform conservation strategies and habitat management decisions. To address this issue, occupancy models are SDMs that rely on repeated visits of spatial sampling units for inferring distribution (\citeproc{ref-mackenzie2017}{MacKenzie et al. 2017}). These models have been widely used in freshwater ecosystems for various taxa (e.g., \citeproc{ref-Wedderburn2022}{Wedderburn et al. 2022}, \citeproc{ref-Couturier2024}{Couturier et al. 2023}).

Second, SDMs rely on the assumption of independent residuals. This assumption may be violated if nearby sampling sites tend to have similar probabilities of species presence, leading to biased estimates of species distribution and potentially inflating the effects of environmental factors (\citeproc{ref-Dormann2007}{Dormann et al. 2007}). Several extensions of occupancy models have been proposed to account for spatial autocorrelation, building on the spatial statistics literature, with conditional autoregressive models (\citeproc{ref-Johnson2013}{Johnson et al. 2013}) and geoadditive models (\citeproc{ref-Rushing2019}{Rushing et al. 2019}). However, these models rely on the Euclidean distance between the spatial sampling units, which does not acknowledge the spatial structure in networks of streams and rivers.

Here I propose spatial occupancy models that account for spatial autocorrelation based on flow connectivity and river network (\citeproc{ref-Peterson2013}{Peterson et al. 2013}). I build on the linear mixed modelling approach developed by Ver Hoef and Peterson (\citeproc{ref-VerHoef2010}{2010}) and Peterson and Hoef (\citeproc{ref-Peterson2010}{2010}), which integrates various distance-based spatial correlation structures (both Euclidean and non-Euclidean) within a single model. I plug-in this variance component approach into occupancy models using a Bayesian approach. A similar approach was recently undertaken by Lu et al. (\citeproc{ref-Lu2024}{2024}) for count data to estimate abundance. Below I outline the statistical developments of this model and demonstrate its application to a semi-aquatic mammal in French streams and rivers.

\section{Methods}\label{methods}

\subsection{Occupancy models}\label{occupancy-models}

To address imperfect detection, I use occupancy models to estimate the true species distribution (\citeproc{ref-mackenzie2017}{MacKenzie et al. 2017}). In these models, monitoring occurs across \(S\) spatial sampling units, or sites. If detection was perfect, the state \(z_i\) of a site \(i\) would be a Bernoulli random variable, taking value 1 with occupancy probability \(\psi_i\) if the site was occupied, and 0 otherwise with probability \(1-\psi\). However, because the ecological process of state occupancy \(z_i\) is only partially observable (since the species might be present but undetected), we must also account for the observation process, which is also modeled as a Bernoulli random variable. When the species is detected at site \(i\), i.e.~\(y_i = 1\), with detection probability \(p\), it confirms that the site is occupied. Conversely, if the species is not detected, i.e.~\(y_i = 0\), with probability \(1-p\), we cannot determine whether the site is occupied or not. Both parameters, \(\psi\) and \(p\), can be modeled as functions of explanatory spatial variables, in the spirit of generalized linear models, and logistic regressions for example.

\subsection{Spatial autocorrelation for stream networks}\label{spatial-autocorrelation-for-stream-networks}

How is spatial autocorrelation accounted for in occupancy models? The usual way is to write the probabilities of occupancy \(\boldsymbol{\psi} = (\psi_1, \ldots, \psi_S)^T\) on some scale, say the logit scale, as a function of \(P\) explanatory variables gathered in a matrix \(\mathbf{X}\) with \(S\) rows and \(P+1\) columns (including a column of 1's for the intercept) with a corresponding vector of regression parameters \(\boldsymbol{\beta} = (\beta_0, \beta_1, \ldots, \beta_P)^T\) that need to be estimated. To account for spatial autocorrelation, a random effect \(\boldsymbol{\epsilon}\) is added to the model, which captures the spatial dependencies among sites (\citeproc{ref-GK2018}{Guélat and Kéry 2018}):

\begin{equation}
\text{logit}(\boldsymbol{\psi}) = \mathbf{X} \boldsymbol{\beta} + \boldsymbol{\epsilon}.
\end{equation}

The random effect \(\boldsymbol{\epsilon}\) can be structured using methods such as conditional autoregressive models and their extensions (\citeproc{ref-Johnson2013}{Johnson et al. 2013}), or geoadditive models (\citeproc{ref-Rushing2019}{Rushing et al. 2019}). However, these approaches typically rely on Euclidean distance to assess proximity among sites, which may not fully capture the complex spatial dependencies present in streams and rivers. Specifically, we are interested in flow connectivity and the topology of streams and rivers (\citeproc{ref-Peterson2013}{Peterson et al. 2013}). Following Ver Hoef and Peterson (\citeproc{ref-VerHoef2010}{2010}) and Peterson and Hoef (\citeproc{ref-Peterson2010}{2010}), I define two sites as flow-connected if water flows from an upstream site to a downstream site, and as flow-unconnected if they share a common confluence downstream but do not directly share flow. Then I parameterize occupancy by rewriting the random effect as a mixture of four components as follows:

\begin{equation}
\text{logit}(\boldsymbol{\psi}) = \mathbf{X} \boldsymbol{\beta} + \boldsymbol{\tau}_{tu} + \boldsymbol{\tau}_{td} + \boldsymbol{\tau}_{eu} + \boldsymbol{\epsilon}
\end{equation}

where \(\boldsymbol{\tau}_{tu}\) is a random effect with spatial covariance between flow-connected sites that can occur in the same direction of the river flow (tail-up), which is suitable for organisms that move passively, such as mussels), \(\boldsymbol{\tau}_{td}\) is a random effect with spatial covariance between flow-connected and flow-unconnected sites, which can occur with or against the direction of flow (tail-down), applicable to organisms that move actively, such as semi-aquatic mammals, \(\boldsymbol{\tau}_{eu}\) is a random effect with a spatial covariance independent of the network topology, influenced by factors like air temperature or precipitation, and \(\boldsymbol{\epsilon}\) is a random effect with variance, often referred to as the nugget, which accounts for additional variability. How to build these covariance components is described in details elsewhere (e.g., \citeproc{ref-VerHoef2019}{Ver Hoef et al. 2019}), and I provide an example in the next section.

\subsection{Case study}\label{case-study}

\begin{figure}[b]

{\centering \includegraphics[width=0.98\linewidth]{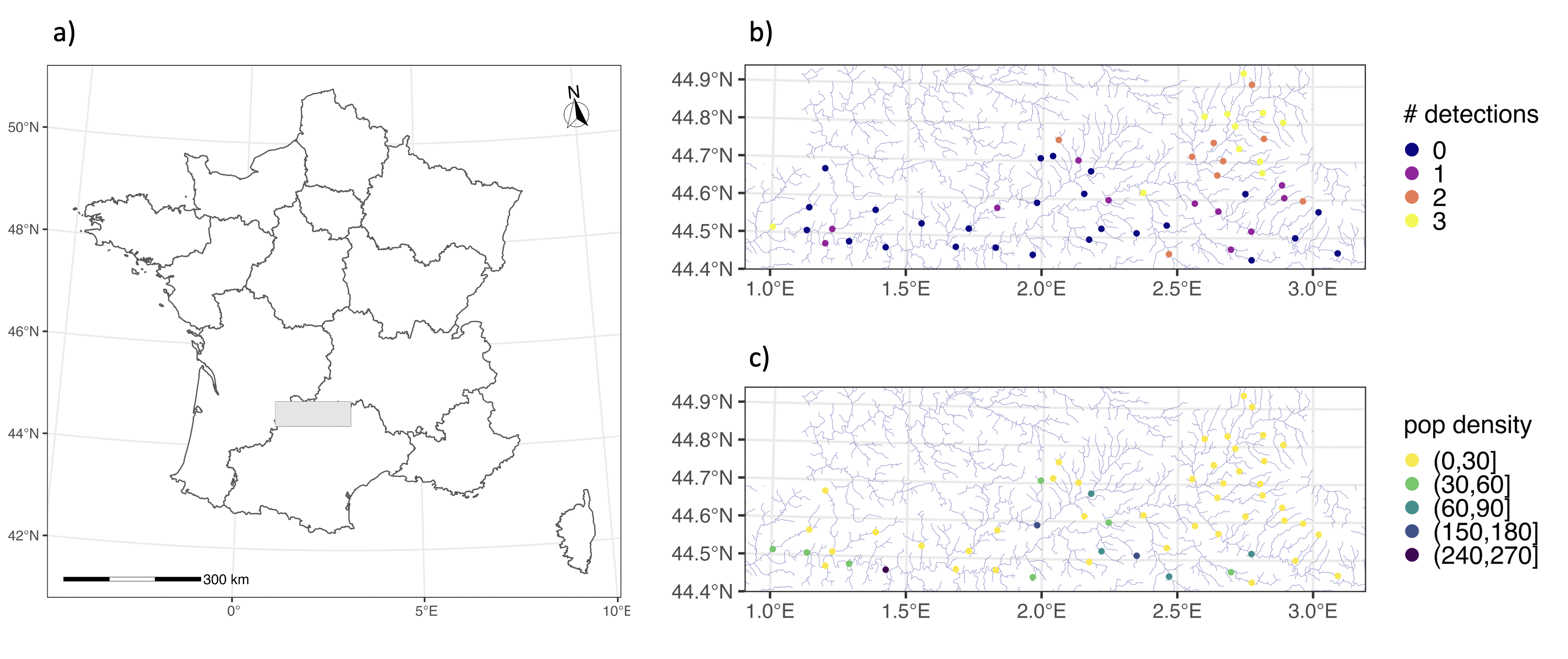} 

}

\caption{Information on the otter data is provided. In panel a), the study area is given in a grey rectable on a map of France. In panel b) the number of detections is given on a map of the study area; it is the sum of detections across the three repeated visits, which can range from 0 (if the species was not detected at any visit) to 3 (if the species was detected at all three visits). In panel c) the human population density is represented on a map of the study area, in number of inhabitants per km2; the missing intervals (90,120] and (120,150] are due to the absence of sites with human population densities within these ranges.}\label{fig:mainfig}
\end{figure}

To illustrate the new approach, I investigated the impact of human disturbance on the occupancy of European otter (\emph{Lutra lutra}) in France. The otter, a semi-aquatic mammal, faced near extinction in the 20th century in France due to extensive hunting for its fur. With hunting bans and protection efforts, the species is now recolonizing the country, and the ecological question is assessing its current distribution. Data on otter detection and non-detection were collected in 2003-2005 in the Midi-Pyrénées region (see panel a in Fig. \ref{fig:mainfig}). Observers searched for signs of otter presence at a small river catchment scale, which was used as the spatial sampling unit. These data were analyzed by Couturier et al. (\citeproc{ref-Couturier2024}{2023}), who found that human density and the proportion of cultivated area influenced occupancy. In this study, I focus on a subsample of this dataset, covering \(S = 56\) sites in the Lot, Aveyron and Cantal counties, which were visited 3 times (see panel b in Fig. \ref{fig:mainfig}). I consided \(P = 2\) explanatory variables. I used human population density as a proxy for human disturbance, calculated as the number of inhabitants per km2 within a 200-m buffer around each stream (see panel c in Fig. \ref{fig:mainfig}). Additionally, I considered the proportion of cultivated areas. Detection was considered as constant. Regarding spatial autocorrelation, since otters can move both downstream and upstream, I used a tail-down model:

\begin{equation}
\text{logit}(\boldsymbol{\psi}) = \mathbf{X} \boldsymbol{\beta} +\boldsymbol{\tau}_{td}
\end{equation}

with an exponential covariance:

\[
    \text{Cov}(\tau_{i},\tau_{j})=
\begin{cases}
    \displaystyle{\sigma^2\exp(-h/\theta)},& \text{if sites } i \text{ and } j \text{ are flow-connected }\\
    \displaystyle{\sigma^2\exp(-(a+b)/\theta)},& \text{if sites } i \text{ and } j \text{ are flow-unconnected }\\
\end{cases}
\]

where I dropped the \(td\) notation in \(\tau_{td}\) for clarity, \(h\) is the stream distance between sites \(i\) and \(j\), \(b\) denotes the longer of the distances between sites \(i\) and \(j\) to the common downstream junction, \(a\) denotes the shorter of the two distances, \(\sigma^2\) is a variance parameter usually referred to as the partial sill and \(\theta\) is a range parameter (see Apendix A in \citeproc{ref-Peterson2010}{Peterson and Hoef 2010}).

\subsection{Simulation study}\label{simulation-study}

I conducted a simulation study to evaluate model performance by examining parameter bias and prediction error. I simulated a stream network over \(S = 100\) sites using an exponential tail-down structure (inspired by the case study, see previous section) with partial sill \(\sigma^2 = 2\) and range parameter \(\theta = 10\). I considered a single covariate, normally distributed with mean 0 and standard deviation 1, with a linear effect (on the logit scale) on the occupancy probability with intercept \(\beta_0 = 0.5\) and slope \(\beta_1 = 1\). I simulated the observation process with a detection probability \(p = 0.6\) across 5 repeated visits per site. I fitted both models, with and without spatial autocorrelation, to the simulated data, and I repeated this procedure 100 times. Eventually, I calculated the relative bias for each parameter and the root mean square prediction error (RMSPE) for each model.

\subsection{Bayesian implementation}\label{bayesian-implementation}

For all analyses, I used the statistical language \texttt{R} (\citeproc{ref-R_Core_Team_2023}{R Core Team 2023}). I used the \texttt{tidyverse} (\citeproc{ref-Wickham_2019}{Wickham et al. 2019}) suite of packages for data manipulation and visualization and \texttt{sf} (\citeproc{ref-Pebesma_2023}{Pebesma and Bivand 2023}) for dealing with spatial data. I used the \texttt{openSTARS} (\citeproc{ref-Kattwinkel2020}{Kattwinkel et al. 2020}) and \texttt{SSN} (\citeproc{ref-VerHoef2014}{Ver Hoef et al. 2014}) packages to build and characterize the network and calculate stream distances. To implement the model, I was inspired by code for spatial stream network models provided in Santos-Fernandez et al. (\citeproc{ref-Santos2022}{2022}) and code for occupancy models from Clark and Altwegg (\citeproc{ref-clark2019}{2019}). I fitted models within the Bayesian framework using the \texttt{rstan} (\citeproc{ref-Stan_Development_Team_2024}{Stan Development Team 2024}) package. I post-processed the results with the \texttt{bayesplot} (\citeproc{ref-Gabry_2019}{Gabry et al. 2019}) package. I specified weakly informative priors for all parameters, specifically normal distributions with mean 0 and standard deviation 1.5 for regression parameters, and uniform distributions for the partial sill, range parameter and detection probability. I ran two chains for a total of 15,000 iterations with a burn-in of 5,000 iterations. I summarized posterior distributions with posterior mean and 95\% credible intervals. I assessed model convergence using R-hat values (\textless{} 1.1), effective sample size (\textgreater{} 100), and visual inspection of the trace plots.

\section{Results and discussion}\label{results-and-discussion}

\begin{figure}[b]

{\centering \includegraphics[width=0.98\linewidth]{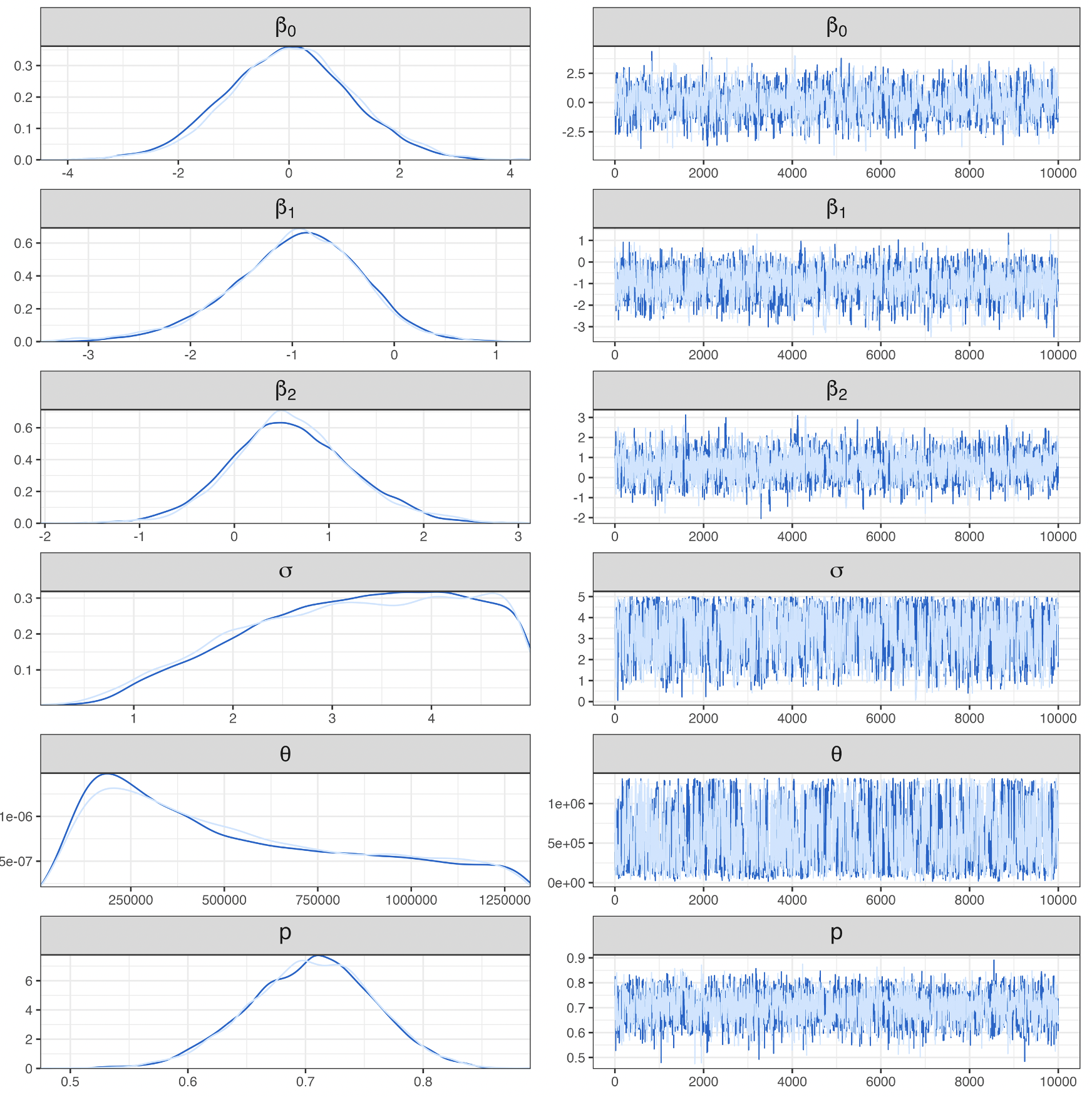} 

}

\caption{Posterior density plots (left) and trace plots (right) are shown for the regression parameters ($\beta_0$: intercept, $\beta_1$: slope of the human population density effect, $\beta_2$: slope of the proportion of cultivated areas effect), spatial parameters ($\sigma$: partial sill, $\theta$: range parameter) and the detection probability ($p$). Each of the two chains is represented in a distinct shade of blue.}\label{fig:postrace}
\end{figure}

Here, I provide the parameter estimates from the new model accommodating spatial autocorrelation (see also Fig. \ref{fig:postrace}), unless otherwise specified. Detection probability was less than one, estimated at 0.71 (0.59, 0.80), which justified the use of occupancy models. The proportion of cultivated area had no effect, with a slope estimated at 0.60 (-0.67, 1.96). Population density also had no effect on occupancy probability, with a slope estimated at -0.96 (-2.24, 0.17). However, I did find a negative effect when spatial autocorrelation was ignored, with a slope estimated at -1.10 (-1.99, -0.34). This latter result aligns with a previous analysis of a more extensive dataset (\citeproc{ref-Couturier2024}{Couturier et al. 2023}) that also ignored spatial autocorrelation.

As anticipated, the effect size of human density increased when spatial autocorrelation was ignored. The most likely explanation for this is a bias due to an omitted variable. There is spatial autocorrelation in human density (see panel c in Fig. \ref{fig:mainfig}), which inflates its effect size; this bias is controlled for when spatial autocorrelation is included in the model. There must be spatial variation in occupancy probabilities attributable to another variable that needs to be accounted for.

The results of the simulation study underscored the importance of accounting for spatial autocorrelation. Ignoring spatial autocorrelation led to a relative bias of -26\% in the slope of the covariate, compared to just 0.7\% when spatial autocorrelation was included. Although the range parameter exhibited substantial bias (330\%), this outcome was expected. In spatial models, there is typically a strong positive relationship between the range parameter and the partial sill (\citeproc{ref-zhang2004}{Zhang 2004}, \citeproc{ref-VerHoef2024}{Ver Hoef et al. 2024}), which showed a bias of 12\%. Importantly, the ratio of the range parameter to the partial sill can be reliably estimated (\citeproc{ref-zhang2004}{Zhang 2004}), as evidenced by its negligible bias (0.9\%). Overall, the root mean square prediction error (RMSPE) was lower for the model with spatial correlation (RMSPE = 0.18) compared to the model without it (RMSPE = 0.23).

Two short-term perspectives arise from this work. From a methodological perspective, the new approach could be extended to multi-season occupancy models, enabling the modeling of colonization probability as a function of distance to habitat features that may impede species movement (e.g., \citeproc{ref-Kervellec2023}{Kervellec et al. 2024}). This would facilitate the quantification of landscape connectivity in freshwater ecosystems. Such development requires moving to spatio-temporal models for stream and river data, which have recently become avaible (\citeproc{ref-Santos2022}{Santos-Fernandez et al. 2022}). From an ecological perspective, the new approach presents significant potential for the analysis of environmental DNA (eDNA). The eDNA methodology offers substantial promise for the non-invasive monitoring of biodiversity in freshwater ecosystems (\citeproc{ref-Carraro2020}{Carraro et al. 2020}). While spatial stream network models have been employed to analyze eDNA data (\citeproc{ref-Winkowski2024}{Winkowski et al. 2024}), these models have overlooked the issue of imperfect detection. Previous studies have recognized occupancy models as effective tools for eDNA data analysis (\citeproc{ref-Burian2021}{Burian et al. 2021}), with some considering spatial autocorrelation (\citeproc{ref-Chen2019}{Chen and Ficetola 2019}), however they have yet to integrate spatial stream networks. The new approach addresses this gap by incorporating both imperfect detection and spatial stream networks.

\section{Acknowledgments}\label{acknowledgments}

First, I would like to warmly thank Jay Ver Hoef and Edgar Santos Fernández for useful discussions on spatial stream network models. Second, I thank Maëlis Kervellec for sharing her code, and Thibaut Couturier for explaining and facilicitating the access to the otter data. Third, I thank all structures and people who participated in the coordination, fieldwork and data management of the otter monitoring. Last, I thank the audience of the GDR2024 and ISEC2024 conferences for their questions about the talk (available at \url{https://doi.org/10.6084/m9.figshare.26968879.v1}) that led to this paper.

\section{Ethics and Integrity statements}\label{ethics-and-integrity-statements}

Not applicable.

\subsection{Data availability statement}\label{data-availability-statement}

Data and code are available at \href{https://github.com/oliviergimenez/spatial-stream-network-occupancy-model}{https://github.com/oliviergimenez/spatial-stream-network-occupancy-model}.

\subsection{Funding statement}\label{funding-statement}

This research is a product of the DISCAR group funded by the French Foundation for Research on Biodiversity (FRB) through its synthesis center CESAB.

\subsection{Conflict of interest disclosure}\label{conflict-of-interest-disclosure}

The author has no conflicts of interest to declare.

\section{References}\label{references}

\phantomsection\label{refs}

\eleft

\clearpage

\newpage

\blandscape

\elandscape

\clearpage

\newpage

\end{document}